# Simulating one hundred entangled atoms using projected-interacting full configuration interaction wavefunctions corrected by projected density functionals

Benjamin G. Janesko

Department of Chemistry & Biochemistry, Texas Christian University, Fort Worth, TX 76129, USA.

b.janesko@tcu.edu

**Abstract**

Simulating entangled atoms is a prerequisite to modeling quantum materials and remains an outstanding challenge for theory. I introduce a correlated wavefunction approach capable of simulating large entangled systems, and demonstrate its application to a 300-electron active space. Projected-interacting full configuration interaction plus density functional theory PiFCI+DFT combines near-exact correlated wavefunctions of multiple partially-interacting model systems, each corrected by a formally exact density functional. This approach can access large active spaces and visualize entanglement and strong correlation while maintaining competitive accuracy for molecular properties.

**Introduction**

Quantum materials exploit quantum entanglement to provide unique properties, including high-temperature superconductivity, beyond-classical measurement sensitivity, and switchable qubits for quantum computing.[1] The textbook example of quantum entanglement is the double-slit experiment (Figure 1a). A single incident electron is "shared" between two slits in a mask, yielding quantum interference in the transmitted electron's probability distribution. ("Sharing" means that the expectation value of each slit's population is not one of the eigenfunctions (0,1,2…) of the population operator.) Measuring one slit's population collapses the wavefunction and eliminates the quantum interference. Many quantum materials share *multiple* electrons among entangled atoms. The diamond nitrogen-vacancy defect (NV$^-$), an optically addressable qubit and quantum sensor, shares six electrons among four entangled atoms surrounding the vacancy.[2] High-temperature superconductivity is hypothesized to arise from strongly correlated electron processes involving large numbers of entangled atoms.[3] Entangled atoms can also occur in chemical reactions. The dissociating hydrogen dimer cation $H_2^+$ shares a single electron among two entangled hydrogen nuclei, a molecular analogue of the double-slit experiment. The dissociating ground-state singlet nitrogen molecule $N_2$ (Figure 1b) shares six electrons among two entangled atoms.

While the double-slit experiment is a staple of undergraduate physics, simulating *multiple-electron* entanglement in quantum materials is a major challenge. The wavefunctions of "normal" molecules and solids are well-approximated by a single molecular orbital configuration (single Slater determinant).[4] The wavefunctions of entangled many-electron systems may instead require a very large number of configurations. For example, a minimal complete active space (CAS) wavefunction of M entangled nitrogen atoms includes on the order of (3M)! configurations: all the ways to distribute 3M bonding electrons among 6M spin-orbitals. Six atoms require more than a mole of configurations! Figure 2a illustrates this imbalance for $N_2$ dissociation. The figure shows dissociation curves computed with four approximate wavefunctions: single-configuration entangled (RHF singlet), ten-configuration entangled (CAS(6,6)SCF singlet), single-configuration unentangled (UHF hextet), and accurate multireference NEVPT2[5]. RHF is qualitatively reasonable for "normal" $N_2$ near equilibrium, but awful at dissociation. Even *computing* the RHF wavefunction is difficult due to variational collapse into artificially stabilized unentangled symmetry broken states.[6] While significant progress has been made, simulating strong correlation remains a significant challenge. State-of-the art GPU-accelerated multireference methods running on classical computers[7] top out at about 100 strongly correlated electrons.[8-9]

Kohn-Sham density functional theory (KS-DFT) provides an alternative. For a system of *non*interacting electrons, entangled or not, the exact wavefunction is a single molecular orbital configuration (apart from some edge cases[10-11]). DFT predicts a real system's ground-state energy and density from a model system of noninteracting electrons, corrected by a formally exact density functional. State-of-the-art DFT codes can treat millions of atoms,[12] raising the prospect of modeling millions of *entangled* atoms. Unfortunately, the exact functional is unknown, and existing approximate functionals struggle to treat strong correlation[13] (*e.g.,* the flat-plane condition[14-16]). The PBE0[17-19] results in Figure 2a exemplify standard DFT. The bond energy near equilibrium is very accurate, but the energy at dissociation is not. Standard approximate functionals also give a "zero-sum" tradeoff between underestimated strong correlation and over-delocalization of charge and spin (Figure SI1).[20] Symmetry breaking can recover accurate energies at dissociation, but lose the entanglement of the reference system wavefunction.[6] Modern flat-plane functionals can be near-exact for model systems at dissociation,[21] but may be unreliable for interacting entangled quantum materials. All of this motivates the present work.

**PiFCI+DFT**

I introduce projected-interacting full configuration interaction plus density functional theory PiFCI+DFT. The thick black curves in Figure 2 show the results. Figure 3 shows the idea. Rather than computing an extremely expensive wavefunction for the interacting-electron system (Figure 3 top), or relying completely on an approximate density functional to correct a noninteracting-electron reference system (Figure 3 middle), I compute and combine *modestly expensive and nearly exact* wavefunctions of multiple projected-interacting reference systems. The approach builds on Hubbard DFT+U[22] and the extension of KS-DFT to partially interacting reference systems.[23-25] I define multiple reference systems, each including an electron-electron interaction projected onto a single state $|\phi_n\rangle$ (eq 1). The bottom panel of Figure 3 illustrates one of these states for $N_{10}$. The real system's exact ground-state energy and density may be obtained from the noninteracting KS-DFT reference system (eq 2), or from any one of the partially interacting reference systems (eq 3), each corrected by a Hartree-exchange-correlation (HXC) density functional. The Hohenberg-Kohn theorems guarantee these functionals' existence.[23] Functional $E^{KS}_{HXC}[\rho]$ corrects the noninteracting KS reference system, projected functional $E^{P,n}_{HXC}[\rho]$ corrects the *n*th partially interacting reference system.[25] Given the exact functionals, any weighted sum of the reference systems (eq 4) *also* gives the exact result. (I assume the exact ground-state density is noninteracting *v*-representable and projected-interacting *v*-representable, *i.e.*, that it is a possible ground-state density of every reference system.) One can choose the states and weights $|\phi_n>,w_n$ so that accurate calculations on the projected-interacting reference systems compensate for the limitations of approximate density functionals.

(1) $\widehat{V_{ee}^{P,n}} = |\phi_n\phi_n> U_n <\phi_n\phi_n|;\ U_n = <\phi_n\phi_n|\widehat{V_{ee}}|\phi_n\phi_n>$

(2) $E_{KS}[\rho] = \min_{\Phi\to\rho} <\Phi|\hat{T}|\Phi> + \int d^3\vec{r}\ v(\vec{r})\rho(\vec{r}) + E_{HXC}^{KS}[\rho]$

(3) $E_{P,n}[\rho] = \min_{\Psi\to\rho} <\Psi|\hat{T} + \widehat{V_{ee}^{P,n}}|\Psi> + \int d^3\vec{r}\ v(\vec{r})\rho(\vec{r}) + E_{HXC}^{P,n}[\rho]$

(4) $E = \min_{\rho} \left\{ E_{KS}[\rho] + \sum_n w_n \left( E_{P,n}[\rho] - E_{KS}[\rho] \right) \right\}$

PiFCI+DFT requires choices. The choices I make here are "black-box", and do not require the user to preselect active spaces, weights, or other system-dependent parameters. I'll discuss each choice in turn.

*Reference System Full CI.* If a system's electron-electron interaction is projected onto a single normalized state (eq 1), the system's full configuration interaction ground-state wavefunction requires at most two Slater determinants.[26] Briefly, starting from the single-determinant reference wavefunction $|\Phi_0>$, separate unitary transforms of the occupied and virtual spaces ensure that only one transformed occupied spinorbital $|\psi_{o\sigma}>$ and one transformed unoccupied spinorbital $|\psi_{v\sigma}>$ have nonzero projections onto $|\phi_n>$. These projections define *occupation numbers* similar to those used in DFT+U: $|<\psi_{o\sigma}|\phi_n>|^2 = n_{p\sigma}$, $|<\psi_{v\sigma}|\phi_n>|^2 = n^v_{p\sigma}$. (If the spinorbitals are expanded in a complete one-electron basis set, then $n_{p\sigma} + n^v_{p\sigma} = 1$.) The bottom panel of Figure 3 illustrates the projection state (black) and transformed orbitals for one of the 40 reference systems used to model $N_{10}$. The reference system full CI wavefunction includes only two Slater determinants $|\Phi_0>$ and $|\Phi_{oo}^{vv}>$. The latter replaces each $|\psi_{o\sigma}>$ in $|\Phi_0>$ with $|\psi_{v\sigma}>$. The off-diagonal element of the 2x2 full CI Hamiltonian is $U_n(n_{p\uparrow}n_{p\downarrow}n^v_{p\uparrow}n^v_{p\downarrow})^{1/2}$. The difference between the two diagonal elements is approximated in terms of the Fock matrix projected onto the transformed orbitals (eq 23 of ref [26]). Table SI1 shows that this approximation has a negligible effect on test molecules.

*Projection States and Weights.* I choose localized atom-centered projection states analogous to the states used in DFT+U. Just as in DFT+U, the results depend on the choice of projection state.[27] For each atom, I project the Kohn-Sham density matrix onto the valence atomic orbitals of the STO-3G minimal basis set. I choose the eigenvectors of this projected atomic density matrix as the projection states. Each projection state defines the electron-electron interaction operator for a different N-electron reference system (eq 1). Projection states on different atoms are typically nonorthogonal. Specialists may note that this is not "nonorthogonal CI": each reference system simply includes a modified electron-electron interaction, and each reference system's wavefunction is computed in the normal way. I choose the weights in eq 4 as the overlap of different projection states $w_n = \sum_m |<\phi_m|\phi_n>|^{-2}$. With this, a projection state orthogonal to all others gets weight 1, and N identical and perfectly overlapping projection states each get weight 1/N. No matter what weights are used, there is no "double-counting" of correlation if the exact functionals are available, because the correlation energy of the *n*th full CI wavefunction and the correction of the *n*th exact density functional sum to zero:

$$(5) \quad 0 = E_{P,n}[\rho] - E_{KS}[\rho]$$
$$= \left( \min_{\Psi \to \rho} <\Psi|\hat{T} + \widehat{V_{ee}^{P,n}}|\Psi> - \min_{\Phi \to \rho} <\Phi|\hat{T}|\Phi> \right) + \left( E_{XC}^{P,n}[\rho] - E_{HXC}^{KS}[\rho] \right)$$

*Orbitals and Orbital Energies.* All calculations use the self-consistent orbitals, orbital energies, and densities from the Kohn-Sham reference system. This choice ensures that all of the partially interacting reference systems have the same uncorrelated electron density. Figure SI1 shows how the choice of KS potential and KS orbital energies affects the partially interacting reference system correlation energies.

*Projected Density Functionals*. I test two approximate density functionals denoted PiFCI+HF and PiFCI+DFT. PiFCI+HF is analogous to CASSCF, and treats all of the XC functionals in eq 2-5 as nonlocal exact (Hartree-Fock) exchange. PiFCI+DFT is analogous to DFT-corrected CASSCF. PiFCI+DFT determines the Kohn-Sham orbitals & orbital energies with the Becke-Lee-Yang-Parr one-parameter global hybrid (25% exact exchange), and models the projected XC functionals as

$$(6) \quad E_{XC}^{KS} + \sum_n w_n \left( E_{XC}^{P,n} - E_{XC}^{KS} \right) = E_{X,HF} + 0.2 \left( E_{X,LDA}^{P,full} - E_{X,HF}^{P,full} \right) + 0.8 E_{C,LYP} + 0.2 E_{C,LDA} + E(D3)$$

(Functionals' density dependence is omitted for conciseness.) This functional uses full exact exchange, LDA and Lee-Yang-Parr correlation,[28] D3 dispersion,[29] and 20% projected LDA exchange in the projection states. Functionals labeled "P,full" use the density matrix projected onto the full set of all projection states.[30] I use a local-hybrid-like *ansatz* to compute the projected LDA exchange energy, eq 15 of ref [31]. Specialists may note that this functional includes 100% full exact exchange in density tails, atomic cores, and all regions outside of the STO-3G valence AO projection states.

*Relation to Other Approximations*. The literature on strong correlation is too large to cover comprehensively. I'll discuss six approximations that are particularly relevant to the present work. (1) CAS-in-DFT approximations use density functionals to add "dynamical" correlation to a multireference CAS wavefunction.[32-33] The related multiconfigurational pair-density functional theory (MCPDFT) computes the full correlation energy from a pair-density functional.[34] The related λ-DFTB and multiconfigurational hybrid methods compute CAS or generalized valence bond wavefunctions for a single reference system employing a globally rescaled electron-electron interaction.[35-36] These methods require an expensive non-black-box CAS calculation or a less expensive black-box valence-bond calculation on the entire active space. PiFCI+DFT, using the choices above, only requires two-determinant CI on the reference systems. PiFCI+DFT can recover a variant of CAS-in-DFT if one chooses a single partially-interacting reference system, in which the electron-electron interaction projected onto the entire active space.[37] (Table SI1 illustrates an example.) PiFCI+DFT can recover a variant of multiconfigurational hybrid methods if one chooses a single partially interacting reference system with $\hat{V}_{ee}^P = \lambda \hat{V}_{ee}$. (2) Selected CI methods compute approximate wavefunctions for large active spaces using deterministic selection (heat-bath CI,[38] DMRG-CAS,[39] DSRG[8]) or stochastic selection (quantum Monte Carlo[40]) of "important" configurations. PiFCI instead selects "important" *projected interactions,* ensuring that each reference system can be treated with full CI. The approaches are complementary: one could imagine a "PiDMRG+DFT" in which the electron-electron interaction is projected onto a large active space, and *e.g.* DMRG-CAS is used to model the projected-interacting reference system wavefunction. (3) Local CI approximations employ localized orbitals to reduce the cost of multiconfigurational calculations.[41] PiFCI+DFT *localizes the interactions, not the orbitals*. The bottom panel of Figure 3 illustrates the difference: for this projected-interacting reference system, the projected electron-electron interaction is spatially localized, and the two transformed orbitals entering the full CI delocalize across the entire system. (4) DFT+U approximations use density functionals to correct the Hubbard model.[22] The key difference is that, while multideterminant treatments of the Hubbard model are available,[42] most modern DFT+U calculations simply use a single-configuration wavefunction and do not compute multireference correlation.[43] (5) Several modern DFT and DFT+U approximations satisfy the flat-plane condition and can treat strong correlation.[14-15, 44-47] PiFCI+DFT grew out of my efforts to derive these methods from correlated wavefunction theory.[25] (6) Many *ab initio* and DFT approaches use symmetry breaking and restoration to "disentangle" entangled atoms.[6, 48] PiFCI+DFT uses entangled, symmetrized wavefunctions throughout, with the hope that these will be useful for capturing entangled-atom physics.

**Results**

Figure 2a shows that PiFCI+DFT accurately treats $N_2$ dissociation. The bond energy near equilibrium is close to the NEVPT2 reference, and the energy at dissociation is almost perfectly zero. The density functional captures the "dynamical" correlation missing from PiFCI+HF, which underbinds near equilibrium. Figure 2b shows that PiFCI+DFT provides a reasonable dissociation curve for $N_{100}$. To my knowledge, this is the first application of a multiconfigurational wavefunction approximation to a 300-electron active space. PiFCI+DFT predicts that this square lattice $N_{100}$ has a relatively weak bond. The equilibrium lattice spacing 2.0 Å is about twice the $N_2$ bond length, and the equilibrium bond energy ~20 kcal/atom is about 15% that predicted for $N_2$. This is chemically reasonable: a nitrogen atom with four nearest-neighbor nitrogens would naïvely have an equilibrium bond energy around ¼ that of $N_2$. PiSCF+DFT correctly predicts that the interaction energy at dissociation is nearly zero: the energy of *entangled* spin-singlet dissociated $N_{100}$ almost perfectly matches the energy of 100 *un*entangled noninteracting nitrogen atoms. Figure SI2 shows results for an evenly spaced chain of ten nitrogen atoms $N_{10}$. PiFCI+DFT predicts an equilibrium lattice spacing 1.5 Å and a bond energy ~25 kcal/atom midway between $N_2$ and $N_{100}$.

PiFCI calculations are fast. Computing the projected full CI correlation energy for 100 nitrogen atoms requires diagonalizing 100 4x4 projected atomic density matrices, then constructing and diagonalizing 400 2x2 CI matrices for the 400 projected-interacting reference systems. A single-shot PiFCI+HF total energy for $N_{100}$ takes less than 30 seconds wall time on a single 3000 MHz AMD EPYC-Milan processor.

PiFCI+DFT provides competitive accuracy for "normal" chemistry. Table 1 reports error statistics for PiFCI+DFT and two standard dispersion-corrected DFT methods, evaluated for subsets of the GMTKN55 benchmark dataset.[49] Table SI2 reports detailed error statistics. PiFCI+DFT provides a weighted mean absolute deviation approaching dispersion-corrected PBE, while also effectively treating entanglement and strong correlation. In particular, PiFCI+DFT provides very low self-interaction error, with a SIE4x4 MAD well below dispersion-corrected B3LYP, consistent with its beyond-zero-sum performance (Figure S1).

PiFCI reference system wavefunctions encode information about entanglement and strong correlation. Table 2 shows that the reference systems' projected occupancy and full CI correlation energy distinguish "normal", entangled, and strongly correlated systems. "Normal" systems like H atom have occupancies near 1. The entangled system $H_2^+$, the molecular analogue of the double-slit experiment, has fractional occupancies near ½. Dissociating symmetry-restricted singlet $H_2$ has occupancies near ½ and large correlation energies. Figure 1c shows this in action. Each nitrogen atom in the predicted equilibrium structure of $N_{100}$ is colored by the full CI correlation energy from its associated reference systems. Values range from small (-0.12 au, blue) to large (-0.19 au, red). The "strong" correlation is strongest for the corner atoms and weakest for the central atoms. Figure 4 illustrates two other model systems, neutral and cationic octane with one dissociating C-C bond. Both bond dissociations produce entangled atoms. The neutral molecule, with two singlet-coupled electrons in the dissociating bond, also displays strong correlation.

**Discussion**

I believe that PiFCI+DFT is the first black-box multiconfigurational wavefunction method capable of routinely treating (at least) 100 entangled atoms and 300 active electrons. I believe that this promising result motivates further development. I'll conclude by discussing possible developments and prospects for the future.

*Self-Consistency*. The current implementation of PiFCI+DFT expands all reference system wavefunctions in a single set of molecular orbitals. This ensures that all reference systems have the same uncorrelated electron density. The first target for future work is self-consistent optimization of these orbitals. I hypothesize that the "strong" correlation captured by PiFCI+DFT will make it easier to converge single-determinant wavefunctions for strongly correlated states, mitigating variational collapse into un-entangled symmetry-broken states. I also note that the exact functionals would ensure that all reference systems the same *correlated* electron density. This could be a useful constraint on approximate projected functionals.

*Solids*. PiFCI+DFT has roots in DFT+U and should be readily extended to periodic systems. Specialists may note that, in DFT+U language, the electron-electron repulsion integral $U_n$ in eq 1 is the nonempirical full (unscreened) U value. PiFCI+DFT does not require choosing or fitting a value of U.[50]

*Properties and Excited States*. The key strength of PiFCI+DFT is that one does not compute the extremely expensive wavefunction for the interacting-electron entangled system. Properties of the real interacting-electron system must be computed as energy derivatives (as is typical in DFT), rather than as expectation values of the real system's wave function. Excited-state properties must be accessed by extending the exact (not linear response) time-dependent Kohn-Sham theory to treat partially interacting reference systems. I hypothesize that this extension will involve excited states of the partially interacting reference systems, just as time-dependent Kohn-Sham theory involves excited states of the noninteracting reference system. Work in this direction is underway.

*Choice of Projected-Interacting Reference Systems*. In this work, I choose to project the electron-electron interaction onto one localized state at a time (eq 1). Each reference system's correlated wavefunction captures (part of) the correlation energy of a singlet-coupled electron pair shared between the reference atom and one or more other atoms. This choice means that the reference systems' correlated wavefunctions do not capture all "important" correlation effects in all systems. While those effects *would be* captured by the exact projected density functionals, the limitations of approximate functionals might warrant different choices of reference system. I'll introduce four examples. (1) The wavefunctions of $C_2$ and stretched $Cr_2$ are not well described by a product of singlet-coupled, shared electron pairs.[51-52] Test calculations (not shown) suggest that the current projected-interacting reference systems do not accurately treat $C_2$ and $Cr_2$. (2) Ionizing an isolated atom typically changes one reference system's occupancy $n_{p\sigma}$ from near 1 to near 0, such that the reference system correlation energy is very small. With these choices, PiFCI+HF atomic ionization potentials are very close to Hartree-Fock theory, thus the PiFCI+DFT ionization potentials in Table 1 are rather inaccurate. (3) Dispersion (van der Waals) interactions arise from coupled fluctuations on pairs of atoms, and cannot be captured one atom at a time. PiFCI+DFT includes D3 dispersion to capture this. (4) Projecting the electron-electron interaction operator onto a single state ensures that none of the reference systems include like-spin correlation.[26] Like-spin "strong" correlation can be important in some real systems.[16] A natural and systematic way to capture all of these effects is to project the electron-electron interaction onto *pairs* of localized states. Work in this direction is underway.

*Choice of Approximate Functionals*. This work uses a DFT "exchange" functional, 20% projected LDA exchange, to capture some dynamical "correlation" effects. The DFT literature includes hundreds of approximate KS functionals, any of which can be converted into approximate projected-interacting functionals using the methods of ref [31]. It remains to be seen which of these approximations will work best. Approximate pair-density functionals developed for MCPDFT[53] might further reduce discrepancies between the wavefunction and DFT pieces (eq 5), though this approach goes outside the Hohenberg-Kohn theorems.

Overall, I believe that these results motivate vigorous exploration of projected-interacting reference systems to treat entangled atoms.

**Materials and Methods**

I have implemented PiFCI+DFT as an extension of the PySCF electronic structure package. The full implementation is freely available online at github.org/bjanesko/CoreProjectedHybrids. Calculations on $N_2$ use the def2-TZVP basis set. Calculations on $N_{10}$ and $N_{100}$ use the def2-TZVP basis set, with Kohn-Sham orbitals obtained from Gaussian 16 DFT calculations using the LDA and the 3-21G basis set. The orbital energies entering the projected-interacting CI are obtained using 25% exact exchange. As discussed above, converging single-determinant self-consistent field calculations on entangled $N_{10}$ and $N_{100}$ was very difficult, as the calculations tended to collapse into artificially stabilized un-entangled states. LDA/3-21G was the only realistic option. Calculations on the GMTKN55 database use the aug-cc-pVQZ basis set with initial guesses from B3LYP/aug-cc-pVQZ calculations.

**Acknowledgments**

I acknowledge the TCU High-Performance Computing Center for computing resources.

**Figures and Tables**

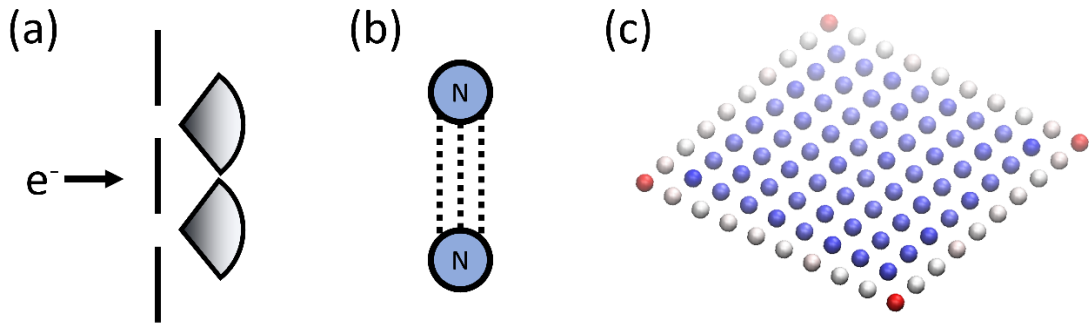

**Figure 1.** Entangled quantum systems. (a) Double-slit experiment. (b) Dissociating nitrogen molecule $N_2$. (c) Dissociating singlet $N_{100}$ as treated here. Atoms in Figure 1c are colored by the magnitude of "strong" correlation, from large (red) to small (blue).

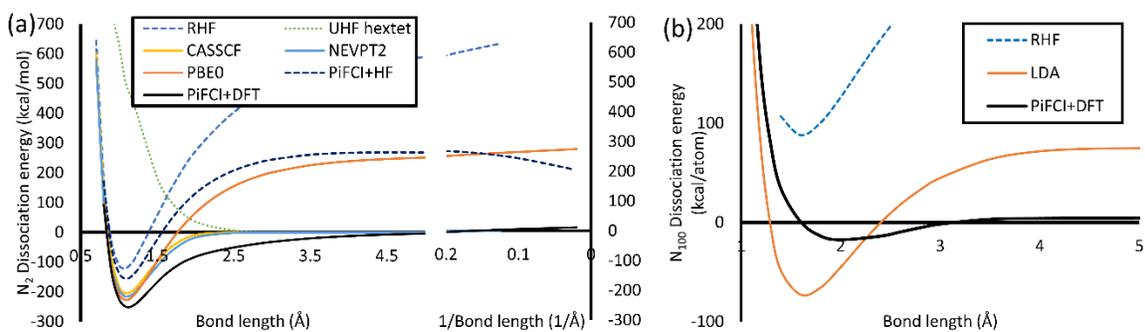

**Figure 2.** Computed potential energy surfaces. (a) Dissociating $N_2$. (b) Dissociating $N_{100}$.

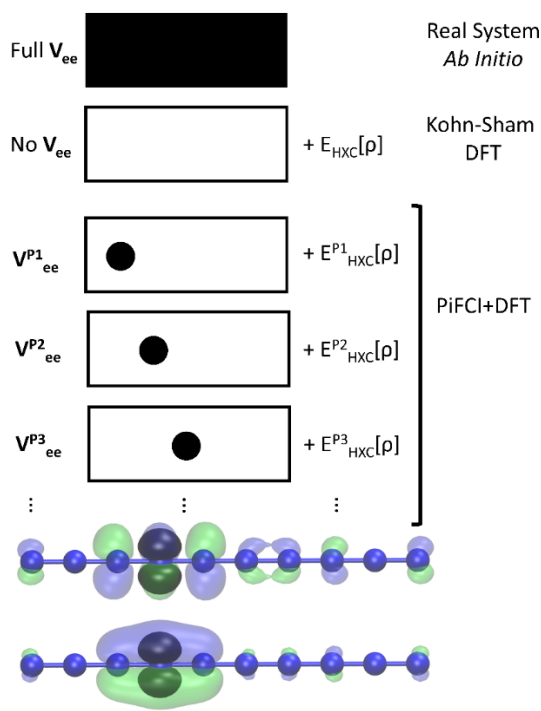

**Figure 3.** Cartoon of PiFCI+DFT. *Ab initio* wavefunction theory treats a real system of fully interacting electrons (top). Kohn-Sham DFT treats a reference system of noninteracting electrons corrected by a density functional. PiFCI+DFT treats multiple reference systems of *partially*-interacting electrons, each corrected by a projected density functional. Given the exact functionals, any combination of the partially-interacting systems' ground-state densities and energies recovers the exact results. The lower panels illustrate one of the 40 reference systems from a PiFCI+DFT calculation on $N_{10}$. The projection state for electron-electron interactions is in black, and the transformed virtual (top) and occupied (bottom) orbitals used in the two-determinant full CI are in blue and orange.

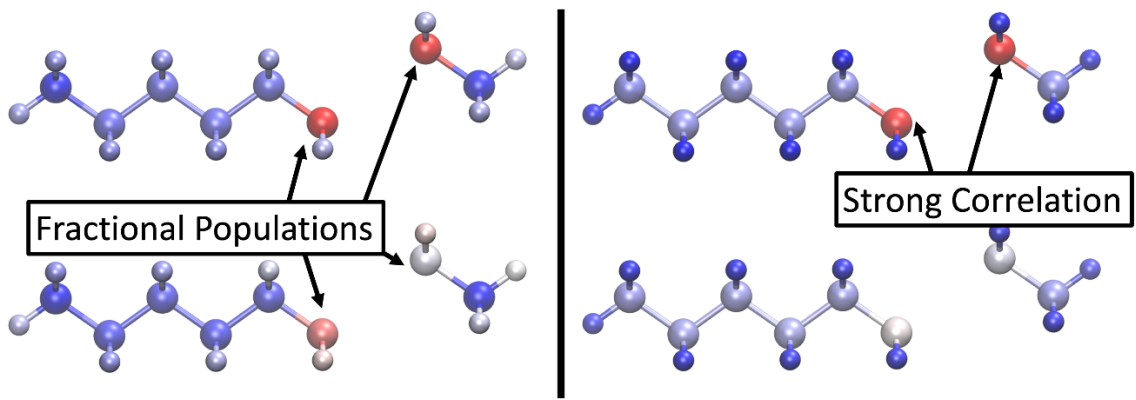

**Figure 4.** Dissociating neutral (top) and cationic (bottom) octane molecule. Each atom is colored by the associated projected-interacting reference systems average occupancy (left) or summed full CI correlation energies (right). Red denotes occupancy near ½ (left) or large correlation energy (right).

| Dataset | PBE-D3(BJ) | B3LYP-D3(BJ) | PiFCI+DFT |
|---|---:|---:|---:|
| Cumulative | 8.36 | 5.05 | 9.66 |
| W4-11 atomization energy | 15.7 | 3.4 | 13.2 |
| G21IP ionization potential | 3.85 | 3.55 | 5.50 |
| SIE 4x4 self-interaction error | 23.72 | 18.06 | 4.83 |
| RSE43 radical stabilization | 2.94 | 1.72 | 1.97 |
| BH76 barrier height | 9.62 | 5.70 | 4.47 |

**Table 1.** Mean absolute errors MAE for representative databases of the GMTKN55 benchmark dataset, and cumulative WTMAD-2 for 45 subdatabases (see SI). B3LYP-D3(BJ) and PBE-D3(BJ) results are from ref [49].

| Classification | System | Population | Correlation energy |
|---|---|---:|---:|
| "Normal" | Isolated H atom | 0.96 | 0 |
| | Equilibrium $H_2^+$ | 0.72 | 0 |
| | Equilibrium $H_2$ | 0.82 | 5 |
| Entangled | Stretched $H_2^+$ | 0.49 | 0 |
| Entangled and strongly correlated | Stretched $H_2$ | 0.46 | 117 |

**Table 2.** Average majority-spin population (unitless) and full CI correlation energy (milli-Hartree) for "normal", entangled, and strongly correlated model systems.

**Supporting Information**

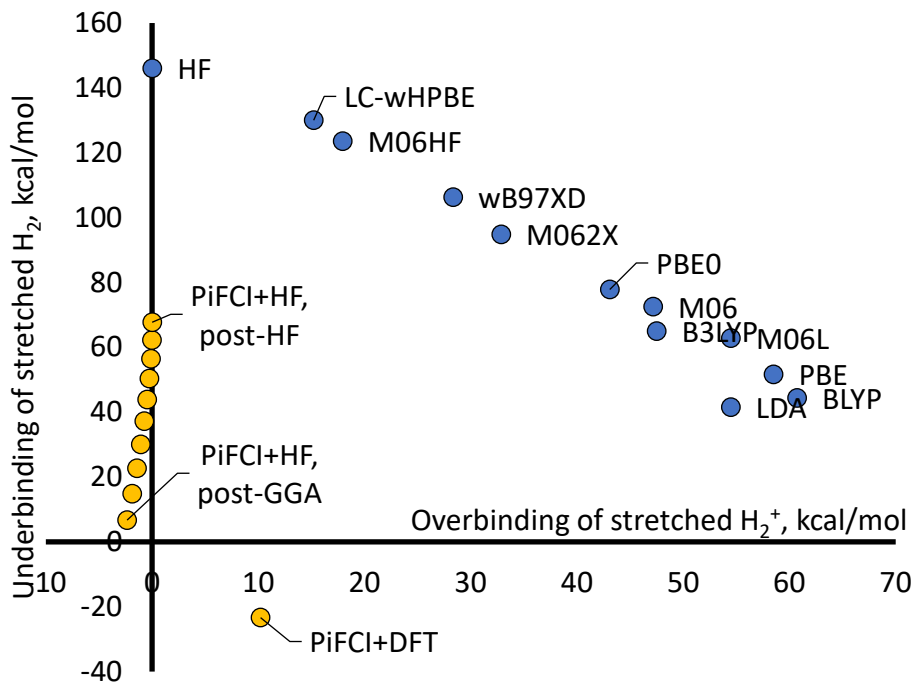

**Fig. S1.** Zero-sum tradeoffs in the bond energy of stretched $H_2^+$ and stretched singlet spin-symmetry-restricted $H_2$. Calculations using the def2-TZVP AO basis set. Standard DFT approximations either over-delocalize $H_2^+$ or underbind $H_2$. PiFCI+DFT provides very accurate results, with modest overbinding of stretched H2. The different points labeled PiFCI+HF points show the effect of changing the Kohn-Sham orbitals and orbital energies entering the reference system CI Hamiltonian. Kohn-Sham orbitals computed using a small (large) fraction of exact exchange give a small (large) HOMO-LUMO gap and correspondingly large (small) reference system correlation energies.

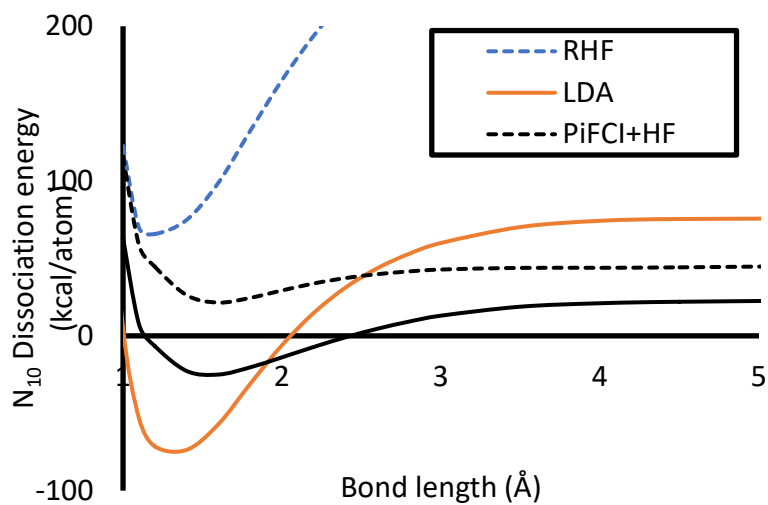

**Fig. S2.** Dissociation curve for linear spin-singlet $N_{10}$ chain.

|  | | | | PiFCI+HF | | |
|---|---|---|---|---|---|---|
| Basis set for MOs | System | Hartree-Fock | Full CI | Approximate diagonal element | Exact diagonal element | Full valence projection |
| STO-3G | H atom | -0.466582 | -0.466582 | -0.466582 | -0.466582 | -0.466582 |
| | $H_2$, $R_{HH}$=0.74 Å | -1.116759 | -1.137284 | -1.121392 | | -1.137284 |
| | $H_2$, $R_{HH}$=1.2 Å | -1.005107 | -1.056741 | -1.031427 | -1.030208 | -1.056741 |
| | $H_2$, $R_{HH}$=10 Å | -0.572320 | -0.933164 | -0.867987 | -0.867987 | -0.933164 |
| | $H_2$, $R_{HH}$=60 Å | -0.550271 | -0.933164 | -0.920333 | -0.920344 | -0.933164 |
| aug-cc-pVTZ | H atom | -0.499821 | -0.499821 | -0.499821 | -0.499821 | -0.499821 |
| | $H_2$, $R_{HH}$=0.74 Å | -1.133034 | -1.172630 | -1.137767 | -1.137285 | -1.153496 |
| | $H_2$, $R_{HH}$=1.2 Å | -1.063040 | -1.112790 | -1.089750 | -1.088698 | -1.110441 |
| | $H_2$, $R_{HH}$=10 Å | -0.741525 | -0.999643 | -0.894788 | -0.893939 | -0.990018 |
| | $H_2$, $R_{HH}$=60 Å | -0.719456 | -0.999642 | -0.897555 | -0.896712 | -0.990457 |

**Table S1.** Total energies (Hartree atomic units) for H atom and $H_2$. PiFCI+HF is exact within a given basis set for H atom. PiFCI+HF using the exactly computed reference system CI Hamiltonian, versus the approximate diagonal element used in the rest of this paper, give nearly identical results. "Full valence projection" denotes PiFCI+HF using a single reference system, in which the electron-electron interaction operator is projected onto the entire set of STO-3G valence AOs. If the MOs are also computed in the STO-3G minimal basis set, PiFCI+HF recovers the exact energy at large $R_{HH}$ and converges as $R_{HH}^{-1}$ towards that limit. Full valence projection recovers full CI at all bond lengths. If the MOs are computed in the larger aug-cc-pVTZ basis set, PiFCI+HF is missing "dynamical" correlation relative to full CI.

| Dataset | Number entries | MD | MAD | RMSD | Max |
|---|---|---|---|---|---|
| W4-11 | 140 | -1.85 | 13.174 | 17.41 | 44.29 |
| G21EA | 24 | -7.62 | 8.788 | 10.76 | 20.98 |
| G21IP | 36 | 0.32 | 5.5 | 7.68 | 22.29 |
| DIPCS10 | 10 | -8.18 | 11.236 | 12.6 | 24.96 |
| PA26 | 26 | 3.56 | 3.908 | 4.78 | 12.49 |
| SIE4x4 | 16 | -2.87 | 4.834 | 5.81 | 10.47 |
| ALKBDE10 | 10 | -10.62 | 11.391 | 13.41 | 21.74 |
| YBDE18 | 18 | -3.4 | 3.943 | 6.26 | 19.78 |
| AL2X6 | 6 | 2.21 | 2.209 | 2.42 | 4.16 |
| HEAVYSB11 | 11 | -4.13 | 5.658 | 6.2 | 8.92 |
| NBPRC | 12 | -1.93 | 4.501 | 5.3 | 8.56 |
| ALK8 | 8 | -6.42 | 6.421 | 9.68 | 19.52 |
| RC21 | 21 | -11.36 | 13.728 | 17.95 | 45.59 |
| G2RC | 25 | 4.65 | 12.723 | 15.93 | 36.91 |
| BH76RC | 30 | 0.75 | 5.115 | 8.89 | 35.97 |
| FH51 | 51 | 5.94 | 8.947 | 13.19 | 47.89 |
| TAUT15 | 15 | 0.45 | 2.801 | 3.51 | 6.82 |
| DC13 | 13 | 3.01 | 32.486 | 43.05 | 85.67 |
| MB16-43 | 43 | -42.34 | 47.68 | 56.6 | 133 |
| DARC | 14 | 32.89 | 32.889 | 33.29 | 43.19 |
| RSE43 | 43 | -0.93 | 1.969 | 2.93 | 12.79 |
| BSR36 | 36 | -3.52 | 3.522 | 4.6 | 13.42 |
| CDIE20 | 20 | -3.03 | 3.382 | 4.09 | 7.24 |
| ISO34 | 34 | 1.01 | 5.683 | 8.88 | 21.84 |
| PArel | 20 | -0.51 | 2.074 | 2.62 | 6.05 |
| BH76 | 76 | -1.35 | 4.469 | 6.47 | 29.82 |
| BHPERI | 26 | 13.36 | 14.685 | 15.33 | 24.6 |
| BHDIV10 | 10 | 4.79 | 6.723 | 9.25 | 20.22 |
| INV24 | 24 | 0.64 | 4.42 | 6.57 | 22.31 |
| BHROT27 | 27 | -0.5 | 0.633 | 1.25 | 4.46 |
| PX13 | 13 | 5.22 | 5.217 | 5.38 | 7.33 |
| WCPT18 | 18 | 7.59 | 7.594 | 7.88 | 11.47 |
| RG18 | 18 | 0.26 | 0.256 | 0.35 | 1.11 |
| ADIM6 | 6 | 0.96 | 0.964 | 1.01 | 1.42 |
| S22 | 22 | 0.01 | 0.576 | 0.66 | 1.47 |
| S66 | 66 | 0.39 | 0.539 | 0.6 | 1.23 |
| HEAVY28 | 27 | -0.25 | 0.402 | 0.52 | 1.39 |
| WATER27 | 27 | 13.38 | 13.383 | 17.4 | 39.9 |
| CARBHB12 | 12 | 0.18 | 0.462 | 0.5 | 0.88 |
| PNICO23 | 23 | 0.58 | 0.651 | 0.8 | 2.33 |
| HAL59 | 59 | -0.16 | 0.643 | 1.02 | 3.86 |
| AHB21 | 21 | -1.41 | 2.007 | 2.3 | 4.35 |
| CHB6 | 6 | -1.66 | 3.132 | 3.64 | 6.51 |
| IL16 | 15 | -0.23 | 0.7 | 0.92 | 2.27 |
| ACONF | 15 | -0.17 | 0.167 | 0.2 | 0.4 |

**Table S2**. PiFCI+DFT error statistics (kcal/mol) for 45 subdatabases of the GMTKN55 database. MD denotes mean deviation, MAD denotes mean absolute deviation, RMSD denotes root-mean-square deviation, Max denotes maximum